\documentclass[11pt]{article}
\pdfoutput=1
\usepackage{jheppub}
\usepackage{epsfig}
\usepackage{amsfonts}
\usepackage{mathrsfs}
\usepackage{latexsym}
\usepackage{color}
\usepackage[usenames,dvipsnames]{xcolor}
\usepackage{amsmath,amssymb}
\usepackage{verbatim}
\definecolor{red}{rgb}{1,0,0}
\numberwithin{equation}{section}
\def\bea{\begin{eqnarray}} 
\def\eea{\end{eqnarray}}
\def\be{\begin{equation}} 
\def\ee{\end{equation}} 
\def\ba{\begin{array}}
\def\ea{\end{array}} 
\def\nn{\nonumber}

\begin{document}

%
%
%

\title{\center Covariant and background independent functional RG flow for the effective average action}

\author[a]{Mahmoud Safari,}
\author[a]{Gian Paolo Vacca}

\affiliation[a]{Dipartimento di Fisica and INFN - Sezione di Bologna,  via Irnerio 46, 40126 Bologna, Italy.}

\emailAdd{safari@bo.infn.it}
\emailAdd{vacca@bo.infn.it}

\abstract{
We extend our prescription for the construction of a covariant and background-independent effective action for scalar quantum field theories to the case
where momentum modes below a certain scale are suppressed by the presence of an infrared regulator. The key step is
an appropriate choice of the infrared cutoff for which the Ward identity, capturing the information from single-field dependence of the ultraviolet action,
continues to be exactly solvable, and therefore, in addition to covariance, manifest background independence of the effective action is guaranteed at any scale.
A practical consequence is that in this framework one can adopt truncations dependent on the single total field.
Furthermore we discuss the necessary and sufficient conditions for the preservation of symmetries along the renormalization group flow. 
}

\maketitle

\section{Introduction}
The functional renormalization group (FRG) approach to quantum field theories is based on the Kadanoff's and Wilson's ideas of coarse-graining
and rescaling~\cite{kadanoff,wk} and is a tool which can give an access to non perturbative physics, as well to the strong field regimes.

One of the formulations~\cite{EAA-Wett,EAA-Morris} deals with the study of the flow of a scale dependent effective action, the generator of the 1PI vertices,
whose definition is based on introducing an infrared (IR) regulator inside the functional integral. 
This object is called the effective average action and formally satisfies an exact flow equation which can be taken as a definition of the quantization of a QFT,
described by the field content and their symmetries. The functional flow describes a trajectory in theory space and in the IR direction points towards the standard effective action,
given an initial condition in the ultraviolet (UV) region.

Effective (average) actions are off-shell objects not directly related to physical quantities and therefore need not be reparametrization invariant,
while S-matrix elements, obtained with the well known LSZ construction, from which physical information is extracted, are invariant under field redefinitions.
While this is not a necessary requirement for an effective action, the property of being a functional scalar in the target field space is definitely more desirable.

The formulation of a geometric approach to field theory, fulfilling these requirements was completed by Vilkovisky and DeWitt~\cite{vilkovisky_gospel1984,vilkovisky_npb1984,dewitt_effact}. The key idea is choosing the quantum field to transform as a vector under field reparameterization~\footnote{See also\cite{FT,BO,Od}.}.
This requires the use of the so called covariant background-field method, where the field in the UV action is split in a nonlinear way into a background and a quantum field.
The implication of the single-field dependence of the UV action for the effective action is encoded in the so called
splitting Ward identity (spWI)~\cite{rebhan_npb1988,hps,burgess_kunstatter,kunstatter_1992,blasi_npb1989}. 
However, the nonlinearity of the quantum-background split in the UV makes the spWI complicated to the extent that it leads to an apparently
separate dependence of the effective action on the background and the average quantum field, so that in the IR manifest background independence is lost.

One can extend this geometric approach to the case of FRG, where the path integration over low moment modes is suppressed by an IR regulator.
The spWI is modified by the cutoff, therefore leading to modified splitting Ward identities (mspWI)~\cite{ pawlowski_review,  pawlowski_0310,  Safari}, and the problem of double-field dependence becomes even more severe in this case. In a non covariant framework, this problem had also been addressed in~\cite{reuter_wetterich_npb1994, litim_pawlowski_0208, morris_1312, morris_1502, Labus:2016lkh}

The study of the flow of the off-shell effective average action adds another motivation for having a covariant approach which is manifestly background independent. Indeed any analysis based on FRG, even if capable of giving non perturbative insights into a QFT, practically requires to work in a submanifold of the theory space, i.e. a truncation, which can still be described by infinitely many couplings.
If there is a dependence in two fields, the background $\varphi$ and the average quantum field $\bar \xi$, then the problem of constructing a reliable truncation becomes very involved and hardly solvable. 
On the other hand if one finds the possibility of describing such a dependence in terms of a single total field, this would highly constrain possible truncations of the effective average action.

Recently, for scalar field theories in the standard approach without a regulator, we have proposed a geometric background-field method leading to an effective action which, while having the benefits of covariance, is free from background dependence and manifestly depends on a single field~\cite{SV}.

In this work we extend our recent results on the construction of a covariant and background independent effective action to the effective average action where an IR regulator is present. After a short review of the mspWI in Section~\ref{SmspWI} and the notion of flat splitting in Section~\ref{Sflatsplit}, in Section~\ref{Sir} we move on to discuss the choice of cutoff which, apart from covariance, allows for a background independent effective average action. Then, in Section~\ref{SRGflow} we study the renormalization group flow equation and show explicitly its covariance and single-field dependence. 
We also derive in the same section the conditions for the preservation of symmetries along the flow. 
We give in this work the basic ideas, with a couple of very simple examples in Section~\ref{Sexamples}.
We then draw some conclusion in Section~\ref{Sconcl}, 
leaving for future works more interesting analysis of models with non trivial symmetries such as $O(N)$ or possibly chiral models,
which might be of interest and have phenomenological implications for the SM physics and beyond.
We have also added an Appendix where we shortly discuss the possibility of extending our arguments to a non exact functional RG framework.

\section{Covariant and single-field effective average action} \label{csfeaa}
For scalar quantum field theories we consider the IR regulated path integral quantization which gives rise to the scale dependent generator of 1PI vertices, known as the effective average action, which turns out to be useful at the non perturbative level. We focus our analysis on this quantity and its Wilsonian RG flow.
For any actual calculation one is forced to adopt the approximation of projecting the flow on a subspace spanned by some families of operators in the so called theory space.
Such truncations must be chosen in the most efficient way according to the symmetries of the problem and can be constrained in several ways.
From one side an approach covariant under field reparameterization is welcome since it helps in removing the redundancy in the operators.
On the other hand, typically for models with non linearly realized symmetries, covariance requires to introduce a background field and therefore one faces the problem of dealing with a possible double field description due to a background-fluctuation split.
Such a dependence is in general constrained by the so called modified splitting Ward identities
which might be solved to restore a single-field description which would be extremely useful to define a truncation.

In this Section we shall start from a solution we have found~\cite{SV} for the splitting Ward identity in the absence of an IR regulator for the standard effective action
and give a prescription to construct a Wilsonian flow which continues to preserve the (unmodified) splitting Ward identities.
The functional RG flow we find is covariant, that is invariant under field reparameterization, so that the effective average action transforms as a scalar.

\subsection{Modified splitting Ward identities}
\label{SmspWI}

In a background field framework for scalar theories, where the field $\phi(\varphi,\xi)$ is split into a background $\varphi$ and a quantum field $\xi$, and in the presence of an infrared regulator which controls the contribution of the fluctuation modes in the path integral, the generator of the connected $n$-point functions $W_k[\varphi,J]$
is a functional of the background field and a source field $J_i$ coupled to the quantum field $\xi^i$. 
Its definition in terms of the ultraviolet action $S[\phi]$ is given by
\be  \label{Wk}
e^{-W_k[\varphi,J]} = \int \!\! D\phi \;\mu(\phi) \,\; e^{-S[\phi]-S_k[\varphi,\xi]- J\cdot\xi},
\ee
where we employ the infrared regulator $S_k[\varphi,\xi] =
{\textstyle{\frac{1}{2}}}\, \xi\!\cdot\! R_k(\varphi) \!\cdot\!\xi$ and with a dot we have denoted integrations as well as internal index contractions.

As usual, on performing a Legendre transform, one defines the IR regulated effective average action~\cite{EAA-Wett,EAA-Morris}, the regulated generator of the 1PI vertices:
\be \label{eaa_def}
\Gamma_k[\varphi,\bar\xi] = W_k[\varphi,J] - J\!\cdot\!\bar\xi -S_k[\varphi,\bar\xi],
\ee
with $\bar \xi=\langle \xi \rangle$, which satisfies the following functional integro-differential equation:
\be
\label{eaa_equation}
e^{-\Gamma_k} = \int \!\! D\phi \;\mu(\phi) \, e^{-S[\phi]+\Gamma_{k; i} (\xi-\bar\xi)^i -S_k[\varphi,\,\xi-\bar \xi] }.
\ee
Here the semicolon ";" denotes a derivative with respect to the quantum field $\xi$ while the comma
will be used for the derivative with respect to the background field $\varphi$ and in general whenever convenient we shall use the deWitt condensed notation.

Taking functional derivatives of this equation with respect to the background fields $\varphi^i$,
we obtain the modified splitting Ward identities (mspWI)\cite{pawlowski_review,  pawlowski_0310, Safari}:
\be
\label{MsWI}
 0  = \Gamma_{,i}+ \Gamma_{;j} \langle \xi^j_{,i}\rangle-{\textstyle{\frac{1}{2}}}\langle\left[ (\xi\!-\!\bar \xi)^m R_{mn} (\xi\!-\!\bar \xi)^n \right]_{,i} \rangle=
\Gamma_{,i}+ \Gamma_{;j} \langle \xi^j_{,i}\rangle-{\textstyle{\frac{1}{2}}} G^{mn}(R_{nm}),_{i} - G^{np} R_{pm}\langle \xi^m\!\!\!,_i \rangle_{;n} 
\ee
where $G^{mn}G_{nl}=\delta^m_l$ with $G_{mn}=\Gamma\!_{;mn}+R_{mn}$ and we have left implicit in the regulator $R_k$ the dependence on the IR scale $k$.
The flow with respect to the RG time $t=\log \frac{k}{k_0}$ of the effective average action $\Gamma_k$ is described by the equation
\be
\label{floweq}
\dot \Gamma_k = {\textstyle{\frac{1}{2}}}\, G^{mn} \dot R_{mn}\,,
\ee
obtained by taking a derivative with respect to $t$ (denoted with a "dot") of Eq.~(\ref{eaa_equation})
and using the property that $\langle(\xi\!-\!\bar \xi)^m (\xi\!-\!\bar \xi)^n  \rangle=G^{mn}$.
For the case of the effective average action it is important to stress that the mspWI 
given in Eq.~(\ref{MsWI}) in its most general form are consistent with the flow,
since one finds \cite{Safari}
\be
\dot {\cal N}_i = -{\textstyle{\frac{1}{2}}} (G \dot R G)^{qp} ({\cal N}_i)_{;pq}.
\ee
where $\mathcal{N}_i$ stands for the expression on the right-hand side of \eqref{MsWI}.
In the following we shall discuss some possible solutions of the mspWI by restricting to a class of quantum-background splits for which the mspWI is exactly solvable. It happens that for these class of splittings the solution is independent of the particular dynamics. 
As we will show, with this choice of quantum-background split, the effective average action $\Gamma_k$ can be written as a function of just a single field $\bar \phi(\varphi,\bar \xi)$
and will be naturally covariant under field reparametrizations.

In order to solve the mspWI we first ignore the regulator terms and solve the spWI for the standard effective action without the infrared regulator. 
Then we shall make a specific choice for the regulator such that 
also the regulator dependent part of the modified splitting Ward Identity given in Eq.~(\ref{MsWI}) identically vanishes.

\subsection{Flat splitting and the covariant single-field dependent effective action}
\label{Sflatsplit}
Here we briefly recall the class of splittings which we have referred to as ``flat quantum-background split'' in~\cite{SV} and, in this case, discuss the solution of the unmodified splitting WI  
\be \label{sWI}
\Gamma,_i + \Gamma_{;k}\,\langle\bar\xi^k\!\!,_i\rangle=0,
\ee
for the effective action $\Gamma$~\cite{SV}. This solution was obtained by requiring the average quantity $\langle \xi^j_{,i}\rangle$ in this equation to be independent of the QFT model described by $\Gamma$, a condition that can be obtained by requiring
the linear dependence  
\be \label{linear_split}
\xi^k\!\!,_i = \alpha^k_i(\varphi) - \beta^k_{ij}(\varphi)\,\xi^j\quad \Rightarrow \quad \langle\xi^k\!\!,_i\rangle = \alpha^k_i(\varphi) - \beta^k_{ij}(\varphi)\,\bar\xi^j = \bar\xi^k\!\!,_i.
\ee
Obviously, $\langle \xi^j_{,i}\rangle$ is a function of $\bar \xi^j$ independent of the dynamics.
As a consequence the splitting Ward identity~(\ref{sWI}) reduces to
\be \label{fsWI} 
\Gamma,_i + \Gamma_{;k}\,\bar\xi^k\!\!,_i=0
\ee
and is integrable if the following Frobenius conditions are satisfied:
\be
\left[\frac{\partial}{\partial \varphi_i} +\bar\xi^k_{,i}\frac{\partial}{\partial \bar \xi_k} \,,\, \frac{\partial}{\partial \varphi_j} +\bar\xi^l_{,j}\frac{\partial}{\partial \bar \xi_l}  \right] =0.
\label{frobenius}
\ee
Given the choice \eqref{linear_split} made for $\bar\xi^k_{,i}$, Eq.~\eqref{frobenius} is equivalent to
\be  \label{int_cond}
d\alpha^k + \beta^k_j\wedge\alpha^j=0, \qquad d\beta^k_j + \beta^k_l\wedge\beta^l_j=0, 
\ee
where $\alpha^k_i$ and $\beta^k_{ij}$ are regarded as tensor valued one-forms and  look like the zero torsion and curvature Cartan structure equations. These are in fact the integrability conditions also for Eq.~\eqref{linear_split}. With the integrability conditions satisfied, any functional $\Gamma[\phi(\varphi,\bar\xi)]$ of the total field $\phi(\varphi,\bar\xi)$, where $\xi(\varphi,\phi)$ is a solution to \eqref{linear_split} solves the spWI.
The general solution to \eqref{int_cond} is given by
\be \label{sol}
\beta^k_{ij} = (U^{-1})^k_a\partial_i U^a_j, \qquad \alpha^k_i = -(U^{-1})^k_a\partial_i f^a,
\ee
which solve the integrability conditions for any $f^a(\varphi)$ and $U^a_i(\varphi)$ as functions of the background field. One then has for Eq.~\eqref{linear_split} the explicit solution
\be \label{xi}
\xi^k(\varphi,\phi) = -(U^{-1}(\varphi))^k_a\, \left(f^a(\varphi) - (g^{-1})^a(\phi)\right),
\ee
where $g$ is an arbitrary function. 
Using this solution the effective action $\Gamma$, which solves Eq.~(\ref{sWI}), can be expressed as a function of the single field $\bar \phi = \phi(\varphi,\bar\xi)$,
which is a function of the background and quantum fields obtained by inverting \eqref{xi}  
\be \label{phi}
\phi^k(\varphi,\xi) = [g\left(f(\varphi) + U(\varphi)\xi\right)]^k.
\ee
Assuming that the functions $f$ and $g$ are invertible, one can choose without loss of generality $g=f^{-1}$. This is equivalent to rewriting the equation in terms of the new background field $\bar\varphi = g(f(\varphi))$ and dropping the bar afterwards, or choosing the boundary condition $\phi^k(\varphi,0)=\varphi^k$. We stick to this choice from now on. 
%
It turns out that equation \eqref{phi} is in fact related to the splitting through an exponential map $[Exp_\varphi\, \xi]_\gamma^i$, with the flat connection
\be
\gamma^k_{ij}= (f^{-1})^k\!\!,_b f^b\!\!,_{ij}  = -(f^{-1}),_{ab}^k\,f^a\!\!\!,_i\,f^b\!\!\!,_j\,.
\label{flatconn}
\ee
Indeed the equality $\phi^i(\varphi,\xi)=[Exp_\varphi\, \xi]_\gamma^i$ holds if we take $U^a_i=\partial_i f^a$, a choice that can be made without loss of generality, as discussed extensively in \cite{SV}.

%

Given the fact that the split is based on an exponential map, the quantum field $\xi^i$ transforms as a vector of the field space, and this guarantees the covariance of the effective (average) action. In fact, furthermore, under a change of coordinates by a function $h$ the quantum field transforms as $\xi\rightarrow \partial h\,\xi$ and the background field transforms as $\varphi\rightarrow h(\varphi)$. If one transforms the function $f\rightarrow f\!\circ\! h^{-1}$, or equivalently the connection $\gamma^k_{ij}$ that we have introduced, Eq.~\eqref{phi} shows that the total field will also transform as $\phi\rightarrow h(\phi)$ as expected. The covariance relation for the effective action is therefore expressed as $\Gamma_{G',f'}[\phi']=\Gamma_{G,f}[\phi]$, where a prime denotes a coordinate transformation. Apart from $f$, the dependence of the effective action on the the tensors appearing in the ultraviolet, collectively denoted as $G$, is made explicit.    
%
Let us also note that, as discussed in~\cite{SV}, the choice of the function $f$ is dictated by the UV symmetries we would like to preserve, so that the presence of the function $f$ cannot be considered as an extra dependence in the effective action $\Gamma$.
\subsection{The choice of an infrared regulator}
\label{Sir}

We will generalize in this section the ideas described above to the case of a scale dependent effective action defined in~\eqref{eaa_equation}.
The presence of the infrared regulator does not affect the covariance of the effective action but modifies the Ward identity \eqref{sWI} to \eqref{MsWI}. 
The simple dependence of the effective action on a single field is violated by the cutoff unless the last two terms of Eq.~\eqref{MsWI} vanish. In the following we introduce a cutoff kernel which fulfills this requirement. Let us therefore focus on the difference among the two identities \eqref{sWI} and \eqref{MsWI}, namely the two cutoff dependent terms in \eqref{MsWI}. The ordinary derivatives in these terms can be replaced with any covariant derivative without affecting the sum~\cite{Safari}.
The difference between the two equations is therefore given by 
\be 
{\textstyle{\frac{1}{2}}} G^{mn}\nabla_i R_{nm} + G^{mp}R_{pn}\langle\nabla_i\xi^n\rangle_{;m}.
\label{diffWI}
\ee
An important point here is that this connection does not have to be torsion free. In fact, if we choose the covariant derivative $\nabla_i = \partial_i + \beta_i$ to be associated to the flat connection $\beta^k_{ij}$, which can in general possess torsion, for flat splitting Eq.~\eqref{diffWI} shows that the second term vanishes by the fact that $\nabla_i \xi^n = \alpha^n_i$ is independent of the fluctuation field. Still the first term does not generally vanish, unless the covariant derivative 
\be 
\nabla_i R_{mn} = (R_{mn}),_i - \beta^k_{im}\,R_{kn} - R_{mk}\,\beta^k_{in}
\ee
is zero. 
It is then not difficult to guess the form of the cutoff kernel whose covariant derivative vanishes. This is 
\be \label{co}
R_{mn} = U^a_m\, \tilde R(-\square)\, U^a_n,
\ee
where $\tilde R$ is a scalar function of the box $\square = \partial_\mu\partial^\mu$, and $U^a_n$ is the matrix appearing in the solution \eqref{sol}. We have been careful about the order of the factors, which is required by the fact that $\tilde R$ is a differential operator. It is easy to verify
{\setlength\arraycolsep{2pt}
\bea 
(R_{mn}),_i &=& (U^a_m),_i\, \tilde R(-\square)\, U^a_n + U^a_m\, \tilde R(-\square)\, (U^a_n),_i \nn\\
&=& (U^{-1})^k_b(U^b_m),_i\, U^a_k\, \tilde R(-\square)\, U^a_n + U^a_m\, \tilde R(-\square)\,U^a_k \,(U^{-1})^k_b (U^b_n),_i \nn\\
&=& (U^{-1})^k_b(U^b_m),_i\, R_{kn} + R_{mk} \,(U^{-1})^k_b (U^b_n),_i \nn\\
&=& \beta^k_{im}\, R_{kn} + R_{mk} \,\beta^k_{in}
\eea}%
which is nothing but $\nabla_i R_{mn}=0$. On the other hand
\be 
\partial_\mu (U^a_n \xi^n) = U^a_n\partial_\mu \xi^n + (\partial_\mu U^a_n) \xi^n = U^a_n\left[\partial_\mu \xi^n + (U^{-1})^n_b(\partial_\mu U^b_m) \xi^m\right] = U^a_n\left[\partial_\mu \xi^n + \beta^n_{\mu m} \xi^m\right] = U^a_n\nabla_\mu \xi^n
\ee
So formally
\be 
\partial_\mu U^a_n = U^a_n \nabla_\mu, \qquad
\nabla_\mu = \partial_\mu + \beta_\mu.
\ee
Using this, one can rewrite the cutoff kernel as
\be 
R_{mn} = U^a_m\, \tilde R(-\square)\, U^a_n = U^a_m U^a_n\, \tilde R(-\nabla^2) = \bar g_{mn}\, \tilde R(-\nabla^2), \qquad \bar g_{mn} \equiv U^a_m U^a_n, \quad \nabla_k\bar g_{mn} =0.
\label{flatmetric}
\ee
With such a choice for the cutoff kernel the splitting Ward identity is not modified and the single-field dependence of the effective action will continue to hold in the presence of the regulator.
Notice that this choice for the cutoff is only dictated by the adopted spitting \eqref{phi} through the $U^n_i$ matrix and has nothing to do with the theory under consideration. We will show in the next section that $U^n_i$ will finally drop out of the flow equation and the results will be as if we had chosen $U^n_i = \partial_i f^n$.

%
%

\subsection{RG flow equation}
\label{SRGflow}

Having discussed the covariance and single-field dependence of the scale-dependent effective action,
we will now move on to the discussion of its flow equation. 
In an expanded form the general formula for the renormalization group flow of the scale-dependent effective action which has been given in Eq.~(\ref{floweq}) reads
\be 
\dot \Gamma\!_{G,f}[\bar\phi] = \frac{1}{2}\,\mathrm{Tr}\left[\left(\Gamma^{(2)}\!_{G,f}[\bar\phi] + R(\varphi)\right)^{\!-1}\!\!\!\dot R(\varphi)\right] ,
\label{CovFlowEq}
\ee
where $\Gamma\!_{G,f}^{(2)}$ is the second fluctuation derivative of the effective average action, and the scale dependencies are suppressed for simplicity of notation.
Similarly to the discussion in~\cite{SV} for the one-loop effective action, one can write  
{\setlength\arraycolsep{2pt}
\bea
(\Gamma\!_{G,f})_{;ij}[\bar\phi]
&=& (\Gamma\!_{G,f}),_p (f^{-1})^p\!\!,_{mn}U^m_{i} U^n_{j} + U^m_{i} (f^{-1})^p\!\!,_m (\Gamma\!_{G,f}),_{pq}  (f^{-1})^q\!\!,_n U^n_{j} \nn\\
&=&
U^m_{i} (f^{-1})^p\!\!,_m \left[(\Gamma\!_{G,f}),_{pq}-\gamma^k_{pq} (\Gamma\!_{G,f}),_k\right] (f^{-1})^q\!\!,_n U^n_j \nn\\
&=&
U^m_{i} (f^{-1})^p\!\!,_m \nabla\!_p\!\nabla\!_q \Gamma\!_{G,f}[\bar\phi] \,(f^{-1})^q\!\!,_n U^n_j,
\label{gamma2}
\eea}%
where the derivatives $(f^{-1})^q\!,_n$ are evaluated at the point $f(\bar\phi)$ and the connection is a function of $\bar\phi$, 
while the matrix $U$ depends on the background field. Using this expression and the cutoff~\eqref{flatmetric} that 
we introduced earlier it is clear that the matrices $U^a_i$ will cancel out in the flow equation~\eqref{CovFlowEq}. 
One can also move the two factors $(f^{-1})^q\!\!,_n$ in~\eqref{gamma2} to the cutoff term so that the flow equation will read
\be 
\dot \Gamma\!_{G,f}[\bar\phi] = \frac{1}{2}\,\mathrm{Tr}\left[\left(\nabla\nabla \,\Gamma\!_{G,f}[\bar\phi] + \hat{R}(\bar\phi)\right)^{\!-1}\!\!\!\dot{\hat{R}}(\bar\phi)\right], 
\label{flow_total_field}
\ee
with the cutoff defined as
\be 
(\hat{R})_{ij}(\bar\phi) = f^a\!\!,_i(\bar\phi)\,\tilde R(-\square)\, f^a\!\!,_j(\bar\phi) \qquad (= f^a\!\!,_i(\bar\phi) f^a\!\!,_j(\bar\phi)\,\tilde R(-\nabla^2), \quad \nabla_\mu \equiv \partial_\mu + \gamma_\mu).
\label{ffcutoff}
\ee
Notice that the term behind $\tilde R$ in the parenthesis above is nothing but the transformation of the metric $\delta_{ij}$ under the change of coordinates $\bar\phi \rightarrow f(\bar\phi)$. Covariance and single-field dependence are manifest in the flow equation~\eqref{flow_total_field}. 

The scale dependent effective action differs from the standard effective action by the presence of the cutoff,
which also appears explicitly in the flow equation~\eqref{flow_total_field}. The explicit form of the cutoff \eqref{ffcutoff} implies that, similarly to the case of the scale-independent effective action discussed in~\cite{SV}, the preservation of the ultraviolet symmetries is subject to the invariance of the connection $\gamma^k_{pq}$. More generally, in the renormalization group context discussed here, one can see that if a symmetry is present in the effective action at some scale, it will be preserved at any other scale, if and only if the flat connection $\gamma^k_{pq}$, used to define the splitting, is invariant under the symmetry.
This can be verified explicitly by checking the invariance of the right hand side of the flow equation~\eqref{flow_total_field}:
suppressing the indices on the effective action $\Gamma$ and taking the transformation $\bar\phi^i \rightarrow \bar\phi'^i$ to be a symmetry $\Gamma'[\bar\phi']=\Gamma[\bar\phi]$ we have 
\be 
\Gamma\!,_{ij}[\bar\phi]-\gamma^k_{ij}(\bar\phi)\, \Gamma\!,_k[\bar\phi] = \phi'^m\!\!\!\!,_i\left[\Gamma'\!,_{mn}[\bar\phi']-\phi^p\!\!,_{m'}\phi^q\!\!,_{n'}(\gamma^l_{pq}(\bar\phi)-\phi'^r\!\!,_{pq}\phi^l\!\!,_{r'})\phi'^k\!\!\!,_l\, \Gamma'\!,_k[\bar\phi']\right]\phi'^n\!\!\!\!,_j,
\label{cond1}
\ee
and also
\be 
f^a\!\!,_i(\bar\phi)\,\tilde R(-\square)\, f^a\!\!,_j(\bar\phi) = \phi'^m\!\!\!,_i[f^a\!\!,_i(\bar\phi')\,\tilde R(-\square)\, f^a\!\!,_j(\bar\phi')]\phi'^n\!\!\!,_j.
\ee
It is clear that the factors $\phi'^m\!\!\!,_i$ and $\phi'^n\!\!\!,_j$ in the two equations above will cancel out in the flow equation~\eqref{flow_total_field}. 
Therefore the invariance 
is subject to the condition 
\be 
\phi^p\!\!,_{m'}\phi^q\!\!,_{n'}(\gamma^l_{pq}(\bar\phi)-\phi'^r\!\!,_{pq}\phi^l\!\!,_{r'})\phi'^k\!\!\!,_l = \gamma^k_{mn}(\bar\phi'),
\label{cond2}
\ee
which is nothing but the invariance of the connection under the aforementioned symmetry. This equation takes its simplest form in the coordinate system where the components of the flat connection vanish. In this case it reduces to $\phi'^r\!\!,_{pq}=0$, which means that in the coordinate system where the  connection vanishes the symmetry transformation must be at most first order in the fields.

In other words, linearizable symmetries, i.e. symmetries that become linear (or more precisely first order in the fields) in some coordinate system~\footnote{As happens for instance when the symmetry group has a fixed point, according to the CWZ lemma~\cite{cwz}.},
can be preserved if one chooses the flat connection to be the one that vanishes in such coordinate system. It is important to notice that linearizability is a statement about the symmetry transformation and moreover theories possessing linearizable symmetries may or may not have a flat field space. 

Summarizing, we have discussed here how to extend the exact RG flow for the effective average action
in order to make it covariant under field reparameterization and single-field dependent. Computing the flow equations, depending on the truncation, can be pretty involved.
There are also other functional RG flow equations, not exact, which nevertheless have beed used giving good results sometimes with less computational efforts. One of them is the Schwinger proper time RG flow.
This can be easily converted in a covariant single-field flow, using the solution of the spWI already given in~\cite{SV}.
We shall briefly discuss this in Appendix A.

\section{Examples}
\label{Sexamples}

According to the discussions of Section~\ref{csfeaa} a covariant and background-independent effective average action is obtained by the exponential split based on any flat connection. The flat connection can be fixed for example by the requirement of preserving linearizable symmetries, therefore leading to a unique effective action in this case\footnote{unless there are fields not charged under the symmetry group, in which case there will be freedom in the choice of the flat connection.}. A theory possessing a linearizable symmetry can have a flat or curved field space. Theories with a curved target space whose symmetries are linearizable provide nontrivial examples where our approach can be fully appreciated. Such theories, in the linear coordinates, may or may not have a good description. In the latter case the theory is considered nonlinear despite the fact that the symmetry is linearizable\footnote{An example of this is the $O(N)$ model on a space with cylindrical topology.}.
However, we need not refer to these coordinates and only take advantage of the fact that the existence of linearly transforming coordinates implies the existence of a flat connection compatible with the symmetries.

Instead of getting involved in such analysis at this point, here in order to illustrate the basic ideas we will be content with two very simple models with flat field space, namely the single-scalar model, and the linear $O(2)$ model on a flat field space.  

Nevertheless, as we just emphasized, this formulation allows to study more general models. In our previous work~\cite{SV} the structure of one-loop divergences for the $O(N)$ model, as a model with linearizable symmetry, was given.
Indeed one can study the flow equations of such theories at some order in the derivative expansion at covariant level, taking advantage of the constraints of single-field dependence, and following the prescription to maintain the symmetries of the model.
It will be interesting to compare the results of this kind of analysis with earlier results on curved linearizable models~\cite{PS,safari_1406}, 
and furthermore look for possible consequences for cases where the symmetries are not fully linearizable~\cite{Codello:2008qq, Percacci:2009fh, Flore:2012ma}.  

%

\subsection{Single-scalar model}

As the first example, let us consider the single-scalar theory in Euclidean space-time, whose most general form of the effective action at the local level and at the second derivative order is
\be 
\Gamma[\phi] = \int  \left[{\textstyle{\frac{1}{2}}}\,J(\phi)\,\partial_\mu\phi\,\partial^\mu\phi + V(\phi)\right].
\label{SFmodel}
\ee
Being one-dimensional, the field space of this model is flat
. We then make the flat splitting $\phi = (f^{-1})\left(f(\varphi)+ \partial f\xi\right)$, which can be considered as a linear splitting followed by a field transformation by $f$.
Since in our approach the flowing effective action is guaranteed to be single-field dependent,
we can start from an expansion around the background and later reconstruct the full dependence.
The second variation of the action with respect the quantum fluctuation $\xi$ at the background ($\xi=0$) can be written as~\cite{SV}
\be
\!\Gamma^{(2)}[\varphi] =\Gamma''[\varphi] - \gamma\, \Gamma'[\varphi] 
=\left[ -J\,\nabla\!_\mu\!\nabla^\mu - \nabla J \partial_\mu\varphi\nabla^\mu
- \nabla J\, \nabla_\mu\partial_\mu\varphi -{\textstyle{\frac{1}{2}}}\nabla^2 J\, \partial_\mu\varphi\partial^\mu\varphi + \nabla^2 V(\varphi) \right]
\ee
where $\nabla$ is the covariant derivative with respect to the connection $\gamma=f''/f'$ as in Eq.~(\ref{flatconn}). For example we have 
\be 
\nabla_\mu = \partial_\mu + \partial_\mu\varphi\, \gamma, \quad \nabla J = J' - 2\gamma J, \quad \nabla^2 J = (\nabla J)' - 3\gamma \nabla J, \quad \nabla V = V', 
\quad \nabla^2 V = V'' - \gamma V'.
\ee 
One can also rewrite it in terms of the covariant derivative $\tilde \nabla$ build with the connection $\tilde \gamma =J'/(2J)$ associated to the ``metric'' $J$,
getting rid in this way of the single derivative operator. Defining 
\be
\delta \gamma=\tilde \gamma-\gamma=\frac{J'}{2J}-\frac{f''}{f'},
\ee
the second variation of the action can be rewritten as
\be
\Gamma^{(2)}[\varphi] = J \left (-\tilde \nabla^2 +Q(\varphi) \right),
\ee
where the function $Q$ reads
\bea
Q(\varphi)&=&\tilde \nabla^2 V+\delta \gamma \,\tilde\nabla_\mu \partial^\mu \varphi+\frac{1}{J} \delta \gamma \,V'\nn\\
&=&\frac{1}{J}\left( V''-\frac{f''}{f'} V'\right)+\left(\frac{J'}{2J}-\frac{f''}{f'}\right) (\partial^2 \varphi+\frac{J'}{2J} \partial_\mu \varphi \partial^\mu \varphi).
\eea
Finally an equivalent expression in terms of normal derivatives is given by
\be \label{gam2}
\Gamma^{(2)}[\varphi]  = -J\partial^2 - J'\partial_\mu\varphi\partial^\mu - {\textstyle{\frac{1}{2}}}J''\partial_\mu\varphi\partial^\mu\varphi + 
{\textstyle{\frac{1}{2}}}\gamma J' \partial_\mu\varphi\partial^\mu\varphi -J' \partial^2 \varphi+ \gamma J \partial^2\varphi + V'' - \gamma V'.
\ee

In order to illustrate the covariance properties it is enough to consider the flow of the potential $V$.
This can be obtained by just considering constant fields in the flow equation,
a fact which drastically simplifies the $\Gamma^{(2)}$ operator in any one of the forms given above. 
Before proceeding let us comment on how one could analyze more general truncations.
For example to extract information at order $O(\partial^2)$ in the derivative expansion one could in principle cast the computation in two ways,
keeping in mind the fact that we can take advantage of the single-field dependence to work mostly at the background level.

The first possible approach is to use directly the flow equation for the action $\Gamma_k$ and compute the trace keeping
the contributions of all the operators present in the truncation to the one of interest, in our case up to $\dot J(\varphi) (\partial \varphi)^2$.
This is a not so trivial task because of the structure of the operator  which is non minimal and contains first order derivatives. 
For general $\varphi$ one can try to expand the denominator in the trace and implement off-diagonal Heat Kernel techniques.

A second, simpler, approach consists in studying the flow of $\Gamma^{(2)}$ in which case to extract $\dot J$ it is enough
to set the background field $\varphi$ to a constant.
The trace has then a bit simpler structure and can be studied in a diagrammatic way using vertices $\Gamma^{(3)}$ and $\Gamma^{(4)}$
also evaluated at constant background field. The details of such computation will be discussed elsewhere.

Instead, here we limit ourselves to 
the local potential approximation (LPA) and therefore take the fields to be constant in the flow equation. In this case space-time covariant derivatives can be taken to coincide with ordinary derivatives.
At the full-field level, this leads to
\be 
\Gamma^{(2)}(\phi) = \left(- J(\phi) \partial^2 + \nabla^2 V(\phi)\right)\;[(f^{-1})'(f(\phi))]^2\,(f'(\varphi))^2
\label{fullgamma2LPA}
\ee
where the last two factors cancel out at $\phi=\varphi$. The potential term is expressed as
\be 
\nabla^2 V(\phi) = V''(\phi) - \gamma\,V'(\phi), \qquad \gamma = -(f^{-1})''(f(\phi))\,(f'(\phi))^2.
\ee
Note that the second variation of the action can also be written as
{\setlength\arraycolsep{2pt}
\bea 
\Gamma^{(2)}(\phi) &=&
\big[-\tilde J(f(\phi)) \;\partial^2 +  \tilde V''(f(\phi))\big]\;\,(f'(\varphi))^2,
\eea}%
where $\tilde J$ and $\tilde V$ are diffeomorphism transformations of $J$ and $V$ by $f$
\be \label{JV}
\tilde J(f(\phi)) =[(f^{-1})'(f(\phi))]^2 J(\phi), \qquad \tilde V(f(\phi)) = V(\phi), 
\ee
so that
\be 
\tilde V''(f(\phi)) = [(f^{-1})'(f(\phi))]^2\, [V''(\phi) +(f^{-1})''(f(\phi))\,(f'(\phi))^2\,V'(\phi)] =[(f^{-1})'(f(\phi))]^2\, \nabla^2 V(\phi).
\ee
According to the prescription of Section~2.4, taking the cutoff to be
\be \label{cutoff}
R_k(\varphi) = (f'(\varphi))^2\,\tilde{R}_k,
\ee 
the flow equation becomes (setting ${\rm Vol}=\int_x$)
\be 
\dot{\tilde{V}}(f(\phi)) = \frac{1}{2{\rm Vol}}\,\mathrm{Tr}\left[\frac{\dot{\tilde{R}}_k}{-\tilde J(f(\phi)) \;\partial^2 + \tilde{R}_k +  \tilde V''(f(\phi))}\right].
\ee
Let us also stress that this expression, even beyond the LPA, can be written both in terms of the ordinary laplacian ($\partial_\mu\partial^\mu$)
or, as in \eqref{flow_total_field}, the covariant one ($\nabla_\mu\nabla^\mu$) for any regular reparameterization. 
The operators inside the trace are simply related by a similarity transformation by $f'(\phi)$ which does not change the eigenvalues of the operators but only the vector space basis.


We stress that the dependence in the background and fluctuating fields $\varphi$ and $\xi$ appears only through the total field $\phi$.
If we had started with a truncation with a non runnning
\be 
J(\phi) = (f'(\phi))^2
\label{LPAcond}
\ee
one would have got a covariantly transformed flow of the LPA truncation of a theory with a trivial kinetic term.
Indeed from the l.h.s equation in \eqref{JV} we would have $\tilde J(f(\phi)) = 1$, and the flow equation would therefore be
\be 
\dot{\tilde{V}}(f(\phi)) = \frac{1}{2{\rm Vol}}\,\mathrm{Tr}\left[\frac{\dot{\tilde{R}}_k}{-\partial^2 + \tilde{R}_k +  \tilde V''(f(\phi))}\right].
\ee
Let us make an important remark here regarding Wilsonian flows. 
We have given a background-field method prescription based on a flat connection whose choice is guided by the requirement of maintaining the symmetries in the flow of the effective action (possibly related to some UV theory). The approach enjoys covariance and background independence. The function $f$ specifying the splitting is always defined in the ultraviolet and never depends on the scale $k$. A scale dependent $f$ would also make the UV action scale dependent through the splitting. On the other hand, in a Wilsonian flow the effective action is scale dependent and in the simple example considered here this means that $J(\phi)=J_k(\phi)$. So the field redefinition that eliminates the $J_k(\phi)$ function would be scale dependent, satisfying $J_k(\phi)=[f_k'(\phi)]^2$. Therefore, beyond the above mentioned approximation, $J_k(\phi)$ cannot be eliminated along the flow. This is also showing how for example the LPA truncation is different from an $O(\partial^2)$ truncation.


A comment is in order here. As Eq.~\eqref{flow_total_field} suggests, the flow equation for general $J(\phi)$ and $V(\phi)$ is related to the standard flow with linear splitting, by a field redefinition $\tilde\phi = f(\phi)$. The approach, however, gives a prescription to construct the flow in any coordinate system. As a consequence, if the standard flow admits a scaling solution $\tilde J^*(\tilde \phi)$ and $\tilde V^*(\tilde \phi)$, this will translate to a scaling solution of the original flow equation given by $J^*(\phi) = (\partial f)^{-2}\tilde J^*(\tilde \phi)$ and $V^*(\phi) = \tilde V^*(\tilde \phi)$. The scaling solutions are therefore in one to one correspondence. Of course within approximations the two scaling solutions correspond to different truncations related by the above mentioned field redefinition. For instance, in three space-time dimensions and at order $\mathcal{O}(\partial^2)$ in a derivative expansion one expects to find both the Gaussian fixed point, which may in general correspond to $J^*(\phi) \neq 1$, as well as the Wilson-Fisher fixed point. 

It may be instructive to consider a more specific example.
Let us set $J=1$ in Eq.~\eqref{SFmodel} and make an exponential reparameterization.
The linear splitting is then transformed into an exponential one $\phi = \varphi \exp(\xi/\varphi)$, which follows from the choice $f(x) = M\log(x/M)$, for some $M$.
An expansion in powers of the quantum field gives
\be
\phi=[Exp_\varphi\, \xi]_\gamma=\varphi+\xi+\frac{1}{2\varphi} \xi^2+\cdots,
\ee
from which one can directly read the flat connection $\gamma(\varphi)=-1/\varphi$.
Changing coordinates according to the reparametrization defined by $f$,
and working with the redefined potential for $\phi$, $ V(\phi)=\tilde V(f(\phi))$
we get for the specific choice of $f$
\be 
\Gamma[\phi] = 
\int \left[{\textstyle{\frac{1}{2}}}\,(M/\phi)^2\partial_\mu\phi\partial^\mu\phi + V(\phi)\right].
\ee
The effective average action in the LPA satisfies the modified splitting WI if we choose the cutoff
\be 
R_k(\varphi) = (M/\varphi)^2 \tilde R_k(-\Box).
\ee
Let us also consider only the background flow, which we know is the same as the full flow. 
Expanding $\Gamma[\varphi \exp(\xi/\varphi)]$ up to second order in $\xi$ 
one easily finds
\be \label{flowV}
\dot{V}(\varphi) = \frac{1}{2{\rm Vol}}\,\mathrm{Tr}\left[\frac{(M/\varphi)^2\dot{\tilde{R}}_k}{-(M/\varphi)^2\,\partial^2 + (M/\varphi)^2 \tilde R_k + V''(\varphi) + (1/\varphi) V'(\varphi)}\right].
\ee
Then defining $ \tilde V(M\log(\varphi/M)) = V(\varphi)$ it is easy to see that the above equation corresponds to the standard LPA flow of $\tilde V$. 
We stress that one can reinsert the full average quantum fluctuation dependence in this relation by substituting $\varphi \to \varphi \exp{(\xi/\varphi)}$.
This means that any derivative of the potential (action) with respect to the quantum average fluctuations $\xi$ has a flow which can be obtained by the background flow \eqref{flowV}.

\subsection{Flat linear $O(2)$ model in Cartesian and polar coordinates}

In this section we consider a scalar model with two degrees of freedom with linearizable $O(2)$ symmetry and a flat field space. In Cartesian coordinates and Euclidean space-time, this is described by the Lagrangian
\be \label{o2c} 
\mathcal{L} = {\textstyle{\frac{1}{2}}}\,\partial_\mu\phi^i\partial^\mu\phi^i + V(\phi^2), \qquad \phi^2 = \phi^i\phi^i.
\ee
We start with performing a standard computation of the flow equation in LPA in this frame. One needs the second derivative of the action 
\be 
\Gamma,_{ij} = \left(-\square + 2V'\right)(P_\perp)_{ij} + \left(-\square + 2V'+ 4\phi^2 V''\right)(P_L)_{ij} \,,
\ee 
where in the last equation we have introduced the projectors
\be
(P_\perp)_{ij} \equiv \delta_{ij} -\phi^i\phi^j/\phi^2, \quad (P_L)_{ij} \equiv \phi^i\phi^j/\phi^2 \,.
\ee
The cutoff kernel is chosen to be $\delta_{ij}\tilde R(-\square)$ which can be written as $\tilde R(-\square)((P_\perp)_{ij}+(P_L)_{ij})$. Then one finds trivially that
{\setlength\arraycolsep{2pt}
\bea
\mathrm{Tr}\left[\left(\Gamma^{(2)}+\tilde R(-\square)\right)^{\!-1}\!\!\dot {\tilde R}(-\square)\right]
&=& \mathrm{Tr}\left[\left(-\square + 2V'\right)^{\!-1}\!\!\dot{\tilde R}(-\square)\right] \nn\\
&+& \mathrm{Tr}\left[\left(-\square + 2V' + 4\phi^2 V''\right)^{\!-1}\!\!\dot{\tilde R}(-\square)\right].
\eea}%
We now repeat the computation in polar coordinates. Let us first define the mapping among the two charts of the target manifold: 
\be \label{cp} 
\phi^1 =\rho \sin\theta =  f^1(\rho,\theta), \qquad \phi^2 = \rho \cos\theta  = f^2(\rho,\theta)\,, \quad
 \partial f = 
 \!\! \left(\ba{cc} \sin\theta & \rho\cos\theta \\ \cos\theta & -\rho\sin\theta \ea\right) ,
\ee
with $[\partial f]_{ij} = \partial_jf^i$, in terms of which \eqref{o2c} takes the form
\be 
\mathcal{L} = {\textstyle{\frac{1}{2}}}\,\partial_\mu\rho\partial^\mu\rho + {\textstyle{\frac{1}{2}}}\rho^2\,\partial_\mu\theta\partial^\mu\theta + V(\rho).
\ee
To find the flow of the effective average potential, according to the general formula, we need to know the connection, which we extract here in an indirect way. 
The non-linear split is constructed according to $\phi^i(\varphi,\xi) = [f^{-1}\left(f(\varphi)+ \partial f\,\xi\right)]^i$, with the function $f^i$ defined in \eqref{cp} and is also given by an exponential map.
Denoting with $\rho_0$ and $\theta_0$ the background fields and with $\xi_\rho$ and $\xi_\theta$ the fluctuations, the explicit form of the split becomes
\bea
\rho&=&\sqrt{(\rho_0+\xi_\rho)^2+\rho_0^2\,\xi_\theta^2}\quad ,
\quad \theta=\arctan{ \frac{\sin \theta_0 \left(\rho_0+\xi_\rho\right)+\rho_0 \cos \theta_0\, \xi_\theta}{\cos \theta_0 \left(\rho_0+\xi_\rho\right)-\rho_0 \sin\theta_0 \,\xi_\theta}}.
\label{polar_split}
\eea
On expanding these expressions in powers of the fluctuations, the coefficients of the quadratic terms are $-1/2$ the connection coefficients:
\bea
\!\!\!\rho&=&\rho_0+\xi_\rho+\frac{1}{2}\rho_0\, \xi_\theta^2-\frac{1}{2}\xi_\rho \xi_\theta^2+\cdots \quad,
\quad \theta=\theta_0+\xi_\theta-\frac{1}{\rho_0} \xi_\rho \xi_\theta+\frac{1}{\rho_0^2}\xi_\rho^2 \xi_\theta-\frac{1}{3}\xi_\theta^3+\cdots,
\eea
so that the only non zero components of the Christoffel symbols are 
$\gamma^\rho_{\theta\theta} = -\rho$ and $\gamma^\theta_{\rho\theta} = 1/\rho$.

We can now compute the second variation of the action in polar coordinates
%
which, at the background level, reads
\be 
\nabla^2 \Gamma
= \left(\ba{cc} 1 & 0 \\ 0 & \rho^2 \ea\right)\left[-\square + \left(\ba{cc} V''(\rho) & 0 \\ 0 &  V'(\rho)/\rho \ea\right)\right].
\ee
The flow of the effective potential is found by computing the following trace, which is given in the same form of Eq.~\eqref{flow_total_field}
%
\be 
{\textstyle{\frac{1}{2}}}\mathrm{Tr}
\left[\left(\ \!\nabla^2 \Gamma+\! (\partial f)^T \tilde R_k(-\square)\partial f\right)^{-1\!}\!(\partial f)^T \dot{\tilde{R}}_k(-\square)\partial f\right].
\label{tr1a}
\ee
%
%
%
Notice that in the above expression everything is a function of the total fields, the argument of $\partial f$ is $(\rho,\theta)$, that is the total fields,
and
\be 
(\partial f)^T \dot{\tilde{R}}_k(-\square)\partial f = (\partial f)^T\partial f \,\tilde{R}_k(-\square) = \left(\ba{cc} 1 & 0 \\ 0 & \rho^2 \ea\right)\tilde{R}_k(-\square),
\ee
since in LPA the connection in the covariant space-time derivative can be set to zero.
Therefore one can rewrite Eq.~\eqref{tr1a}  as
\be 
{\textstyle{\frac{1}{2}}}\mathrm{Tr} \left\{\left[-\square + \tilde R_k(-\square) + V''(\rho) \right]^{-1}\!\! \dot{\tilde{R}}_k(-\square) \right\}\!+\! 
{\textstyle{\frac{1}{2}}}\mathrm{Tr} \left\{\left[-\square + \tilde R_k(-\square) +  V'(\rho)/\rho \right]^{\!-1} \!\!\dot{\tilde{R}}_k(-\square)\right\}\!.
\label{floweffO2}
\ee
This matches the result found in LPA in the Cartesian coordinates, using the relation $V(\rho) = \tilde V(\rho^2)$, $\rho^2 = \phi^i\phi^i$ where the potential in the Cartesian case is renamed to $\tilde V$. 
This shows the covariance.

Finally, let us make a brief comment on the extension of this analysis to the case of $N$ scalars in a curved field space. In an $O(N)$ invariant model of $N$ scalars, the most general form the kinetic term can take in the Cartesian, or linearly transforming, coordinates is 
\be 
\left[Z(\phi^2)\delta^{ij} +Y(\phi^2)\phi^i\phi^j\right]\partial_\mu\phi^i\partial^\mu\phi^j.
\ee
Although the metric behind the kinetic term is curved, according to our prescription the connection that defines the exponential map used for the splitting must be the one that vanishes in this Cartesian coordinates, and is therefore not compatible with the metric above. The connection then transforms accordingly when the parametrization is changed. In other words, independent of the functions $Z$ and $Y$, the exponential split must be chosen such that in this coordinate system it reduces to a linear split. In general, the functions $Z$ and $Y$ may include inverse powers of the field which is a sign that the theory is nonlinear (in this case has the topology of a cylinder), and the theory must be described in polar coordinates. Our prescription applies to these cases as well.

%
%

\subsection{Comments on the dimensionless flow}

The framework that we have developed so far results in a covariant and single-field dependent flow for the effective average action, which is induced by the coarse-graining procedure. The flow equation can be used to construct within some approximation scheme the standard quantum effective action (at $k=0$). 

A separate point to be addressed is the process of rescaling, which is the final step in the renormalization program. 
If our flow equation continues to be covariant after rescaling 
then the critical properties of the two equations are trivially guaranteed to be the same. 
But, while the ``Trace'' part of the flow equation is still covariant (see \eqref{flow_total_field}), the covariance of the fixed point condition is not obvious at first sight. 
In fact the definition of the fixed point depends on our choice of parametrization. 
More explicitly, for two parametrizations $\phi$ and $\phi'$, denoting with a ``tilde'' the quantities after (the $k$-dependent procedure of) rescaling, we \nolinebreak have
\be \label{fpc} 
\partial_t \tilde{\Gamma}(\tilde{\phi}) - \partial_t \tilde{\Gamma}'(\tilde{\phi}') = \partial_t \tilde{\phi}^{\prime i} \;\partial_{\tilde{\phi}^{\prime i}} \tilde{\Gamma}(\tilde{\phi}) -  \partial_t \tilde{\phi}^i \;\partial_{\tilde{\phi}^i} \tilde{\Gamma}(\tilde{\phi}). 
\ee
So the condition $\partial_t \tilde{\Gamma}(\tilde{\phi})=0$ is not necessarily the same as $\partial_t \tilde{\Gamma}'(\tilde{\phi}')=0$. One faces two possible situations in this case depending on whether or not the field redefinition relating $\phi$ and $\phi'$ is scale covariant, at the quantum level. 
Consider a general reparametrization $\phi = h(\phi')$. Scale covariance of the field redefinition implies that the same relation 
holds between the rescaled fields $\tilde\phi = h(\tilde\phi')$. This implies that
\be 
\partial_t \tilde{\phi}^i \;\partial_{\tilde{\phi}^i} \tilde{\Gamma}(\tilde{\phi}) = \partial_t \tilde{\phi}^{\prime i}  \partial_{\tilde{\phi}^{\prime i}}h^k(\tilde{\phi}')\;\partial_{\tilde{\phi}^k} \tilde{\Gamma}'(\tilde{\phi}') = \partial_t \tilde{\phi}^{\prime i} \;\partial_{\tilde{\phi}^{\prime i}} \tilde{\Gamma}(\tilde{\phi}),
\ee
which forces the equivalence of the fixed point conditions $\partial_t \tilde{\Gamma}(\tilde{\phi}) = \partial_t \tilde{\Gamma}'(\tilde{\phi}')$. An example of such a field redefinition is the relation between Cartesian and polar coordinates in the $O(N)$ model, or the relation between polar coordinates with different parametrizations of the angular fields. This of course poses no constraint on how dimensionless fields can enter in the field redefinition, but instead restricts the ways the dimensionful fields can appear. For instance, in \eqref{cp} or even the split \eqref{polar_split}, dimensionful fields are related in a homogeneous way. The detailed analysis of such models is left for a future work.


Instead, for a field redefinition that violates scale covariance, $\partial_t \tilde{\Gamma}(\tilde{\phi}) = \partial_t \tilde{\Gamma}'(\tilde{\phi}')$ does not necessarily hold. However such field redefinitions involve a dimensionful parameter which makes the r.h.s of \eqref{fpc} vanish in the infrared. To see this more clearly, take as an example a scalar theory with one degree of freedom and consider a general nonlinear field redefinition of the form $\phi = M\,\bar h(\phi'/M)$, where $\bar h(x) = x + \mathcal{O}(x^2)$ and $M$ is a parameter with the dimension of the field. Denoting the wave-function renormalizations by $Z$ and defining the rescaled fields as $\tilde\phi=Z^{1/2} k^{1-d/2} \phi$ and $\tilde\phi'=Z^{1/2} k^{1-d/2} \phi'$, the dimensionless version of the field redefinition will be $\tilde\phi = \tilde M\,\bar h(\tilde\phi'/\tilde M)$ where $\tilde M = Z^{1/2} k^{1-d/2} M$. For $d>2$ it is now clear that in the infrared limit where $\tilde M \rightarrow \infty$ the rescaled fields will become equal $\tilde\phi = \tilde \phi'$, and the r.h.s of \eqref{fpc} vanishes. To be more concrete, consider the flow equation for the potential in this scalar theory in LPA$^\prime$ after the field redefinition. In terms of the rescaled field this is given, using \eqref{flow_total_field}, by  
\be \label{dimlessflow}
\partial_t \tilde{V}'(\tilde{\phi}') + d \tilde{V}'(\tilde{\phi}') - \frac{d\!-\!2\!+\!\eta}{2}\,\tilde\phi'\, \tilde{V}^{(1)}(\tilde{\phi}) = \frac{1}{2}\mathrm{Tr}\left[\frac{(2\!-\!\eta)r + \partial_t r}{-\square/k^2 + r+ \nabla^2 \tilde V(\tilde\phi')/[\tilde h^{(1)}(\tilde\phi')]^2}\right],
\ee
where the second covariant derivative of the potential is
\be 
\nabla^2 \tilde V(\tilde\phi') = \tilde V^{(2)}(\tilde\phi')-\tilde h^{(2)}(\tilde\phi')\tilde V^{(1)}(\tilde\phi')/\tilde h^{(1)}(\tilde\phi').
\ee
Here we have taken the cutoff $\tilde R$ introduced in \eqref{co} to be of the form $\tilde R = Z k^2 r(-\square/k^2)$ with $\partial_tZ/Z \equiv -\eta$ and defined $\tilde\phi = \tilde M\,\bar h(\tilde\phi'/\tilde M) \equiv \tilde h(\tilde\phi')$. It is obvious that in the infrared $\tilde h$ tends to the identity function and the flow \eqref{dimlessflow} reduces to the standard flow equation without the background field method.

The dimensionful scale $M$ might be interpreted typically as a UV mass scale, since one has in mind to redefine the fields in the bare action. Its appearance is therefore nothing really new and is exactly what happens when introducing a UV scale for the bare action. As usual, in order to study the critical scaling properties of the flow one has to remove such a scale by sending it to infinity. In other words, if the redefinition of the field is done at the ultraviolet scale $\Lambda$, where the bare action is defined,  then in the limit $\Lambda \to \infty$, this difference in the definition of the scaling behavior is removed.

To summarize, we have discussed the covariance of the scaling equation under field redefinitions, 
which implies that the fixed point and the critical exponents of a theory and its field redefined version are the same. 
Some interesting cases include scale covariant field redefinitions in sigma models enjoying some symmetry, as in the $O(N)$ models mentioned above.


\section{Conclusions}
\label{Sconcl}
The desirable feature of having a covariant effective action is realized in the geometric formulation of Vilkovisky and DeWitt. 
This approach is based on a covariant quantum-background split by the exponential map, whose definition depends on a connection on field space.
In what is called Vilkovisky's unique effective action, this connection is chosen to be the one compatible with the metric of the theory. 

While having the advantage of preserving the symmetries of the theory in the IR, for theories with a curved target space this quantization procedure
turns the single-field dependence of the UV action into a double-field dependence in the effective action,
which contrary to the spirit of the background-field method, does not allow for a reconstruction of the effective action from the background result. 
This formulation can be extended to the case of the (scale dependent) effective average action, 
and the problem of double-field dependence will show up as an obstacle in finding suitable truncations. 

In our recent analysis we have taken advantage of the idea that the requirement of covariance is realized by any connection
and not necessarily the one compatible with the metric defining the theory. 
We have exploited this fact to find quantum-background splits for which the spWI is significantly simplified and an exact solution can be found,
implying that the single-field dependence of the effective action is manifest. 
These are quantum-background splits based on the exponential map with a flat connection, and lead to a quantum-background dependence in the effective action which is independent of the specific QFT. 

Here, in the context of FRG we have shown that by a suitable choice of the IR regulator the mspWI will continue to be exactly solvable. 
We have therefore given a prescription for an effective average action and its exact renormalization group flow 
which possesses the two desirable ingredients of covariance and background independence at the same time. 
Moreover, we have argued that in theories where linearizable symmetries are present, the flat connection is uniquely fixed
by the requirement that the symmetries be preserved along the renormalization group flow.
This is done by choosing the connection to vanish in the linear coordinates.

It might be worth mentioning that nonlinearizable symmetries may become linear by adding extra degrees of freedom to the theory,
as happens for example for the $O(N)$ symmetry of a theory defined on the $(N\!-\!1)$--sphere, which becomes linearizable by extending the space to a cylinder $\mathbb{R}\times S^{N-1}$.
We have illustrated some simple features of the construction by giving two rather trivial examples with flat field spaces.
We plan to perform in a future work a detailed renormalization group study of the nontrivial case of a field theory on curved field space
with linearizable symmetry, such as the $O(N)$ invariant theory of $N$ scalars, where the full power of the method will be demonstrated.

The problem of background independence for scalar field theory has also been addressed in other works, such as \cite{morris_1312}. 
and especially in the context of conformally reduced gravity in \cite{morris_1502} and \cite{Labus:2016lkh}. 
In these works a background-field approach with linear split is adopted and the main goal is to solve along with the flow equation the constraint coming from the mspWI, 
which becomes a nontrivial task in the presence of the cutoff. 
Here instead, we are concerned with the covariance of the background-field method, which in the sense of Vilkovisky and DeWitt is in conflict with single-field dependence 
even in the absence of the cutoff, and our aim has been to solve the mspWI while keeping the covariance of the approach. 

Our current analysis does not apply to gauge theories, that involves a gauge fixing procedure which tipically introduces an explicit background field dependence
in the off-shell effective action. We plan to return to this problem elsewhere.


\appendix
\section{Non exact RG:  proper time flow}
We have setup a framework where it is possible to analyze in principle exact RG flow of the effective average action keeping covariance under field parameterization and single-field dependence.
The price to pay is that the computation of the trace for truncations going beyond the Local potential approximation,
that is with non trivial kinetic terms requires more efforts.
We would like to comment here on the possibility of using an alternative not exact functional RG scheme, the so called Schwinger proper time RG scheme,
which has been applied to study several models with pretty good
results~\cite{Liao:1994fp, Floreanini:1995aj, Bonanno:2000yp, Mazza:2001bp, Litim:2001hk, Bonanno:2004sy, Litim:2010tt},
for example to study linear $O(N)$ models at criticality.
Since even the exact RG approach can be only used in an approximated way by introducing truncations, it might be worth to consider also this alternative possibility.
Indeed the construction of an approximated RG flow within a proper time approach is simpler, and some interesting results can be obtained using directly standard Heat-Kernel methods.

Starting from the one loop correction to the effective action (with an infinite subtraction), represented in terms of the Schwinger proper time integral by expressions as
\be 
{\rm Tr} \log{[\Omega/\Omega_0]}= - \int_0^\infty \frac{{\rm d}s}{s} \left(e^{-s\, \Omega}-e^{-s\, \Omega_0}\right),
\ee
a proper time flow is naturally defined by
\be \label{ptf}
\partial_t \Gamma= -\frac{1}{2}{\rm Tr}  \int_0^\infty \frac{{\rm d}s}{s} \partial_t \rho_t(s) e^{-s \,\Gamma^{(2)} },
\ee
for a suitable regulator $\rho_t(s)$ which depends both on the "RG-time" $t=\log{k/k_0}$  and the proper time $s$. 
A commonly employed choice for it has been the family of regulators
\be
\rho_t (s,m)=\frac{ \Gamma(m+1,Z s k^2)-\Gamma(m+1,Z s \Lambda^2) } {\Gamma(m+1)},
\ee
which depends on an integer parameter $m$ (often considered in the infinity limit) and on a function $Z$ which may be associated to the truncation of the action.
Plugging the scale derivative of the regulator  
\be
\partial_t \rho_t (s,m)=-\frac{2}{\Gamma(m+1)} (Z s k^2)^{m+1} e^{-Z s k^2}
\ee
in \eqref{ptf}, one finds 
\be
\partial_t \Gamma={\rm Tr}\left[\left( \frac{Z k^2}{\Gamma^{(2)} + Z k^2}\right)^{m+1}\right].
\ee
Our approach for computing at one loop in perturbation theory the covariant single-field effective action~\cite{SV}
can then be directly extended in this context to write a flow equation according to the proper time scheme
which by integration along the "RG-time" gives a resummed formula. Guided by the requirement of covariance and single-field dependence
one can consider the regulator $\rho_t$ such that $Z$ has the same behaviour as $R_k$, being proportional to the induced metric $\bar{g}_{mn}$ associated to the flat connection, as in Eq.~(\ref{flatmetric}).
In this way the structure of the regulator allows to keep the covariance of $\partial_t \Gamma$, while the single-field dependence is assured by the flatness of the connection.


\end{document}